\documentclass[lettersize,journal]{IEEEtran}
\usepackage{amsmath,amsfonts}
\usepackage{algorithmic}
\usepackage{algorithm}
\usepackage{array}
\usepackage[caption=false,font=normalsize,labelfont=sf,textfont=sf]{subfig}
\usepackage{textcomp}
\usepackage{stfloats}
\usepackage{url}
\usepackage{verbatim}
\usepackage{graphicx}
\usepackage{cite}
\usepackage{xcolor}
\usepackage{multicol}
\usepackage{float}
\usepackage{siunitx}

\definecolor{darkgreen}{rgb}{0.0, 0.5, 0.0}
\definecolor{darkorange}{rgb}{0.5, 0.3, 0.1}
\definecolor{orange}{rgb}{1,0.5,0}

\begin{document}

\title{Variance Based Transmitter Localization in Vessel-Like Molecular Communication Channels}

\author{Dağhan Erdönmez, H. Birkan Yilmaz%
\thanks{All authors are with the Department of Computer Engineering, Bogazici University, 34342 Istanbul, Turkiye.

Emails: daghanerdonmez@gmail.com; birkan.yilmaz@bogazici.edu.tr.}

\thanks{This work has been submitted to the IEEE for possible publication. Copyright may be transferred without notice, after which this version may no longer be accessible.}
}

\maketitle

\begin{abstract}
Transmitter localization in vessel-like molecular communication channels is a fundamental problem with potential applications in healthcare. Existing analytical solutions either assume knowledge of emission time or require multiple closely spaced receivers, which limits their applicability in realistic scenarios. In this letter, we propose a simple closed-form approximation that exploits the temporal variance of the received molecular signal to estimate the distance between the transmitter and the receiver without emission time information. The method is derived from a Gaussian approximation of the received signal near its peak and gives an explicit variance-distance relation. Simulation results in physically relevant capillary vessel scale show that the proposed method achieves distance prediction with error on the order of $1\%$.
\end{abstract}

\begin{IEEEkeywords}
Molecular communication, diffusion, localization, blood vessels, variance method.
\end{IEEEkeywords}

\section{Introduction}
Molecular communication via diffusion (MCvD) is a rapidly developing paradigm in nanonetworking, enabling communication between biological or synthetic nanomachines in fluidic environments. Potential healthcare and biomedical applications include abnormal cell activity detection, health monitoring, and targeted drug delivery~\cite{farsad2016survey, varshney2019abnormality, Nakano_Eckford_Haraguchi_2013, etemadi2022abnormality, felicetti2014molecular}. 

A fundamental problem in MCvD systems is the localization of the emission source, i.e., transmitter (TX), since identifying the origin of molecular signals may correspond to locating diseased cells or a drug-release site. Recently, analytical formulations tailored for vessel-like environments (VLEs) under Poiseuille flow have been proposed in the literature \cite{VLE_localization}.

However, existing analytical solutions are subject to restrictive assumptions. The work in \cite{VLE_localization} shows that a single receiver (RX) suffices if the emission start time is known, but in the more practical case where the emission time is unknown, at least two receivers are required. Both conditions limit the applicability of these methods in realistic biomedical scenarios, where emission time is difficult to know and deploying multiple closely spaced receivers may not be feasible. 

In this letter, we propose an analytical localization method that overcomes these limitations. The proposed approach, VAriance-based LOcalization and Ranging (VALOR), exploits the temporal variance of the received molecular signal to infer the TX–RX distance. It requires only a single receiver and does not assume knowledge of the emission time. By deriving an explicit variance–distance relation based on a Gaussian approximation near the received signal peak, we obtain a computationally efficient estimator. Particle-based simulations in capillary-scale environments are used for validating the proposed approach, demonstrating distance prediction accuracy with error on the order of 1\%.
\begin{figure}[!t]
    \centering
    \includegraphics[width=1.0\linewidth]{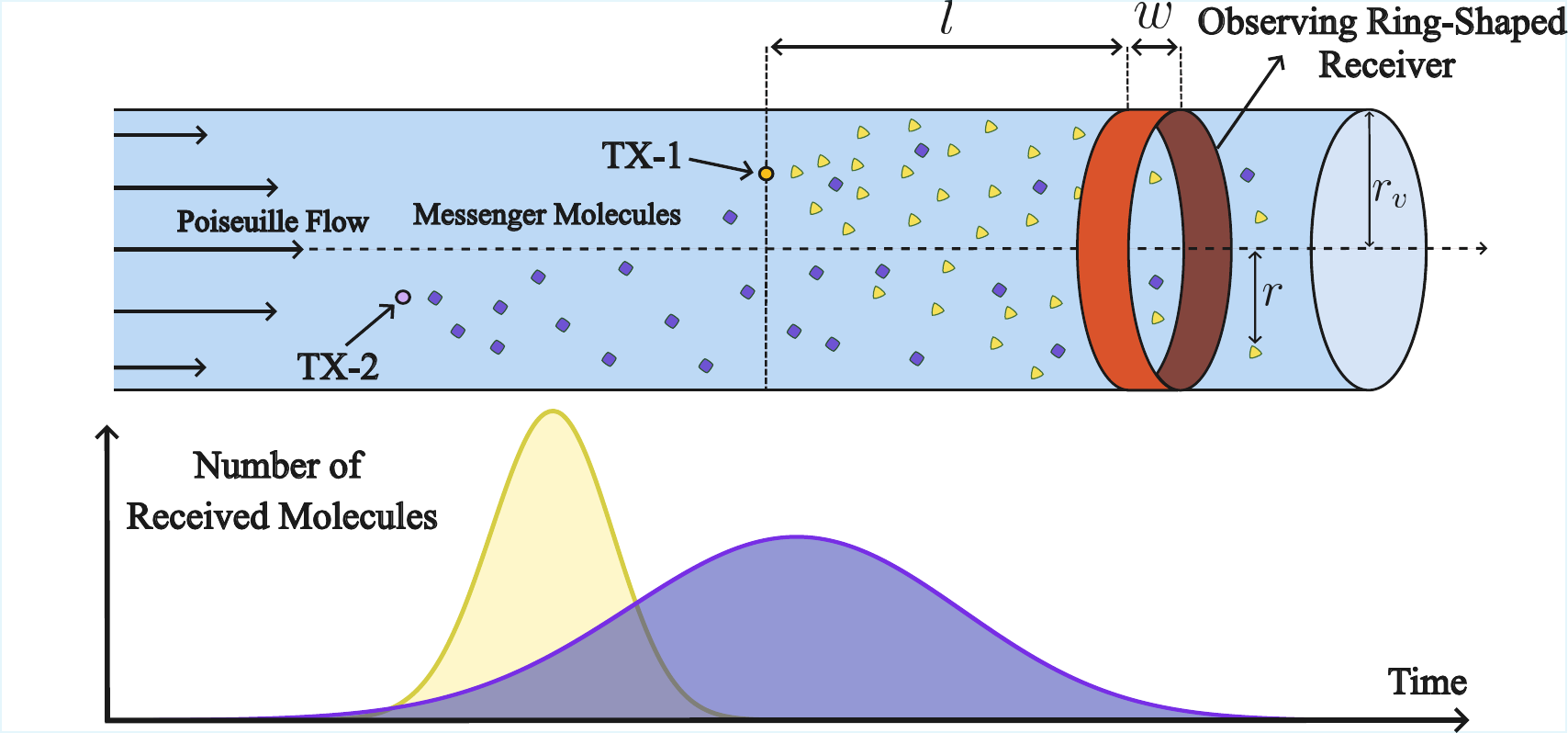}
    \caption{Vessel-like MCvD channel with ring-shaped observing receiver}
    \label{fig:channel}
\end{figure}

\section{System Model and Problem Definition}
In this letter, we consider a cylindrical (vessel-like) environment with fully reflective lateral boundaries and unbounded ends, where particles released from a point emitter propagate by diffusion and advection under a Poiseuille flow profile. A ring-shaped observing receiver is placed downstream on the vessel boundary, sharing the same radius $r_v$ and spanning a width $w$ (Fig.~\ref{fig:channel}). 
The receiver detects all particles that enter its projection volume.

\subsection{Propagation Model}
The net displacement of a particle within a diffusive–advective medium over a single time step $\Delta t$ can be expressed as
\begin{align} \label{delta_r}
    \Delta \vec{r} = 
    \bigl(
        \Delta X_{\textit{diffusion}} + \Delta X_{\textit{flow}},\;
        \Delta Y_{\textit{diffusion}},\;
        \Delta Z_{\textit{diffusion}}
    \bigr),
\end{align}
where the random displacements due to diffusion are
\begin{align}
    \Delta X_{\textit{diffusion}}, \Delta Y_{\textit{diffusion}}, \Delta Z_{\textit{diffusion}} &\sim \mathcal{N}(0, 2D\Delta t).
\end{align}

The deterministic displacement due to flow acts along the axial direction of the vessel and is given by
\begin{align}\label{delta_x}
    \Delta X_{\textit{flow}} = v(r)\Delta t 
    = 2v_{\textit{avg}}\!\left(1 - \frac{r^2}{r_v^2}\right)\!\Delta t,
\end{align}
where \(v(r)\) denotes the parabolic Poiseuille velocity profile \cite{bruus2007theoretical}, 
\(v_{\textit{avg}}\) is the average flow velocity, 
\(D\) is the diffusion coefficient, 
\(r_v\) is the vessel radius, 
and \(r\) is the radial distance of the particle from the central axis. In this work, we align the flow with the \(x\)-axis, chosen as the central axis.

\subsection{Channel Model}
The Péclet number quantifies the relative importance of advection to diffusion in particle transport and is defined as
\begin{align}
    \text{Pe} = \frac{\text{advection transport rate}}{\text{diffusive transport rate}} 
    = \frac{v_{\textit{avg}} r_v}{D}.
\end{align}  

The validity of the model presented below requires 
\begin{align}\label{condition1}
    \text{Pe} \ll 4\frac{l}{r_v},
\end{align}
with \(l\) denoting the distance between the TX and the RX.

The probability for a molecule to be at a certain distance \(l\) from the emission point at time \(t\) is written as \cite{probstein2003physicochemical}
\begin{align}\label{p(l,t)}
    p(l,t) = \frac{1}{\sqrt{4\pi D_e t}}\exp{\left(-\frac{(l-v_{\textit{avg}} t)^2}{4D_e t}\right)}\text{,}
\end{align}
where 
$
    v_{\textit{avg}} = \frac{v_{\textit{max}}+v_{\textit{min}}}{2}
$
is the average flow velocity of the Poiseuille profile across the vessel cross-section, and
$
    D_e = D(1+\frac{1}{48}\text{Pe}^2)
$
is the effective diffusion coefficient. Here, the introduction of the effective diffusion coefficient embeds the influence of the cylindrical vessel geometry into the model, thereby transforming the original three-dimensional transport problem into an equivalent one-dimensional form. The accuracy of this reduction depends on the validity of condition~(\ref{condition1}).

To find the probability of a single molecule being detected by a receiver with width \(w\) at distance \(l\) at time \(t\), we take the integral of (\ref{p(l,t)}) from \(x=l\) to \(x=l+w\). 
\begin{align}
    \textbf{P}(l,t)=\int_{l}^{l+w} p(l',t) dl'
\end{align}

That is, the probability that the molecule lies between \(x=l\) and \(x=(l + w)\) at time \(t\). For small \(w\) this integral becomes
\begin{align}\label{small_w_approx}
    \int_{l}^{l+w}p(l',t) dl' \approx wp(l,t)\text{,}
\end{align}
so
\begin{align}\label{P(l,t)}
    \textbf{P}(l,t) \approx wp(l,t) = \frac{w}{\sqrt{4\pi D_e t}}\exp{\left(-\frac{(l-v_{\textit{avg}}t)^2}{4D_e t}\right)}.
\end{align}

An important observation is that although $\textbf{P}(l,t)$ is Gaussian with respect to the spatial variable $l$, it is \emph{not} Gaussian with respect to time $t$. For a fixed receiver position (e.g., $l = l_1$), the function $\textbf{P}(l_1,t)$ characterizes the probability distribution of molecule arrival times at that receiver. In the case of many molecules, this distribution closely corresponds to the overall shape of the received signal pattern.

\subsection{Problem Definition}
The received signal at the RX is a stochastic counting process.
For $M$ released molecules, the expected number of molecules observed at time $t$
is given by $\mathbb{E}[N_{\mathrm{rx}}(t)] = M\,\mathbf{P}(l,t)$.
The actual received signal is modeled as
\begin{align}
    N_{\mathrm{rx}}(t) = M\,\mathbf{P}(l,t) + \epsilon(t),
\end{align}
where $\epsilon(t)$ represents random fluctuations due to diffusion and molecular
counting noise, whose variance depends on the signal amplitude.

Our objective is to estimate the TX-RX distance $l$ from the
observed signal $N_{\mathrm{rx}}(t)$ using a single receiver, without knowledge
of the emission start time. The TX and RX are not synchronized, so the received signal is subject to an unknown time offset $\tau_{\text{offset}}$.

Formally, the problem is to construct an estimator
\begin{align}
    \hat{l} = \Psi_{\boldsymbol{\Omega}}\!\big(N_{\mathrm{rx}}(t+\tau_{\text{offset}})\big),
\end{align}
where $\Psi_{\boldsymbol{\Omega}}(\cdot)$ denotes a distance estimation function
parameterized by $\boldsymbol{\Omega}=\{v_{\textit{avg}}, D_e\}$.

\section{VAriance-based LOcalization and Ranging (VALOR)}
\begin{figure*}
    \centering
    \includegraphics[width=1\linewidth]{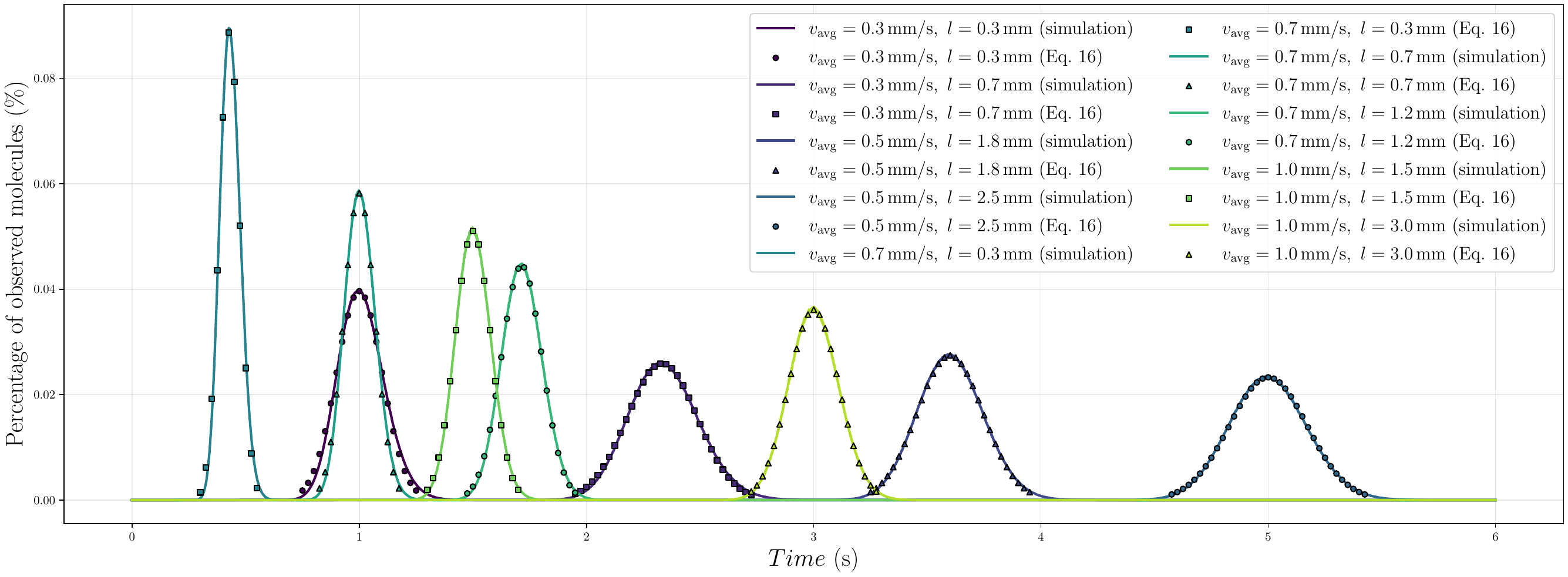}
    \caption{Comparison of simulation results and the Gaussian approximation in (\ref{P_approx}) for the percentage of sensed molecules under different environment settings. ($D = 300 \mu m^2/s, r_v = 5\mu m, w = 1\mu m$)}
    \label{fig:channelmodelapprox}
\end{figure*}

We observed that the temporal variance of the received signal is closely related to the distance that we aim to estimate (see Fig. \ref{fig:channelmodelapprox}). To obtain a closed-form relation between the received signal variance and the TX–RX distance, we consider the exponent term of \(\textbf{P}(l,t)\),
\begin{align}
    E(t) = -\frac{(l-v_{\textit{avg}}t)^2}{4D_e t}.
\end{align}

This function attains its maximum at 
\begin{align}
    t_{\textit{peak}} = \frac{l}{v_{\textit{avg}}},
\end{align}
which roughly corresponds to the time at which the receiver observes the peak molecular signal, ignoring the minor effect of the $\sqrt{t}$ term in the denominator of \eqref{P(l,t)}.

To characterize the local behavior of \(\textbf{P}(l,t)\) near \(t_{\textit{peak}}\), we expand \(E(t)\) as a second-order Taylor series around \(t_{\textit{peak}}\):
\begin{align}
    E(t) \!\approx\! E(t_{\textit{peak}}) 
    \!+\! E'(t_{\textit{peak}})(t\!-\!t_{\textit{peak}}) 
    \!+\! \tfrac{1}{2}E''(t_{\textit{peak}})(t\!-\!t_{\textit{peak}})^2
\end{align}

Since both \(E(t_{\textit{peak}})\) and \(E'(t_{\textit{peak}})\) vanish, only the quadratic term remains, yielding
\begin{align}
    E(t) \approx -\frac{(t-t_{\textit{peak}})^2}{2\left(\tfrac{2D_e l}{v_{\textit{avg}}^3}\right)}.
\end{align}

Substituting this approximation into \(\textbf{P}(l,t)\)  in \eqref{P(l,t)} gives
\begin{align}\label{P_approx}
    \textbf{P}(l,t) \approx 
    w \sqrt{\frac{v_{\textit{avg}}}{4\pi D_e l}}
    \exp{\left(-\frac{(t-t_{\textit{peak}})^2}{2(\tfrac{2D_e l}{v_{\textit{avg}}^3})}\right)},
\end{align}
where the arrival pattern is approximately Gaussian in time with variance
\begin{align}\label{varianceapprox}
    \sigma^2 = \frac{2 D_e l}{v_{\textit{avg}}^3}.
\end{align}

Note that while \eqref{P(l,t)} is Gaussian in space (\(l\)) but not in time (\(t\)), this local approximation reveals that around \(t_{\textit{peak}}\), the received signal can be treated as approximately Gaussian in time. This time-domain Gaussian behavior reflects what the receiver actually observes, enabling us to exploit its variance for distance estimation.

Hence, the TX–RX distance can be estimated directly from the measured temporal variance of the received signal as
\begin{align}
    \hat{l} = \frac{\hat{\sigma}^2 v_{\textit{avg}}^3}{2 D_e}.
\end{align}

This provides a simple closed-form estimator that does not require knowledge of the emission time and relies only on the observed waveform variance, $\hat{\sigma}^2$, at the receiver.

\subsection{Validity of the Gaussian Approximation}
The Gaussian approximation relies on truncating the Taylor expansion of $E(t)$ around 
$t_{\textit{peak}}$ after the quadratic term, so that the received signal is locally Gaussian in time. 
To assess whether this truncation is justified, we examine the higher-order derivatives at $t_{\textit{peak}}$ and evaluate the magnitude of the cubic and quartic terms when $(t-t_{\textit{peak}})$ is on the order of $\sigma$, i.e., within the characteristic width of the Gaussian.

This leads to the following dimensionless correction factors:
$\alpha_3 = \bigl|E^{(3)}(t_{\textit{peak}})\bigr| \,\sigma^3,$
$\alpha_4 = \bigl|E^{(4)}(t_{\textit{peak}})\bigr| \,\sigma^4$.
By explicit calculation we reach 
\begin{align}
    \alpha_3 = 3\sqrt{2}\,\sqrt{\frac{D_e}{l v_{\textit{avg}}}},\; & \;
    \alpha_4 = \frac{24 D_e}{l v_{\textit{avg}}}.
\end{align}

These quantities measure the relative size of the cubic and quartic terms compared to the quadratic 
one at the natural scale of the Gaussian. The approximation is therefore accurate provided that both 
$\alpha_3 \ll 1$ and $\alpha_4 \ll 1$, which can be summarized compactly as
\begin{align}\label{condition2}
    \frac{D_e}{l v_{\textit{avg}}} \ll 1.
\end{align}

It is worth noting that this condition is complementary to the diffusion-dominated requirement at \eqref{condition1}. While the earlier model favors large $D$, small $v$, and large $l$, our Gaussian approximation is more accurate when $D_e$ is small, $v_{\textit{avg}}$ and $l$ are large. The only consistent requirement is thus a sufficiently large TX–RX separation $l$. 

\section{Performance Evaluation}
To test the claims, we used a particle-based Monte Carlo simulator. The simulator explicitly follows the trajectory of each messenger molecule by updating its displacement at every simulation interval according to (\ref{delta_r})-(\ref{delta_x}). Based on these outcomes, we evaluate our Gaussian approximation approach under a variety of system settings. Unless otherwise noted, the number of released molecules is set to $M = 10^6$, and simulation results are reported as the average over 1000 independent realizations. The physical environment parameters are adopted from \cite{VLE_localization, mohrman2018cardiovascular}, with vessel radius typically between \SI{5}{\micro\meter} and \SI{10}{\micro\meter}, consistent with capillary-scale biological environments. The simulation time step is fixed at $\Delta t = 0.1\text{ms}$.

\subsection{Verification of the Channel Model Approximation}
We first verify the accuracy of the derived Gaussian approximation in (\ref{P_approx}) by comparing it with particle-based simulations. 
As shown in Fig.~\ref{fig:channelmodelapprox}, the analytical expression closely matches the simulation, confirming the validity of the approximation in the considered parameter regime.

\subsection{Performance of the Variance-based Localization}
\begin{figure}[t]
    \centering
    \includegraphics[width=1\linewidth]{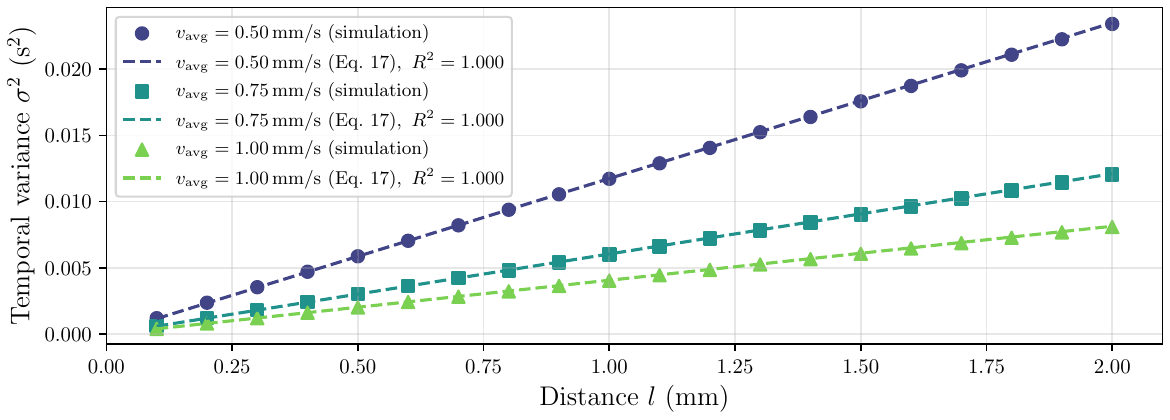}
    \caption{Temporal variance of the RX signal versus distance for different velocities. Scatter points show simulation results, while dashed lines show the theoretical prediction $\sigma^2 = \tfrac{2D_e}{v^3}l$. Our proposed model's coefficient of determination values (denoted as R$^2$) indicate the agreement between simulation and \eqref{varianceapprox}. ($D = 300 \mu m^2/s, r_v = 5\mu m, w = 1\mu m$)} 
    \label{fig:varianceapproxplotforv}
\end{figure}

\begin{figure}[t]
    \centering
    \setlength{\abovecaptionskip}{0pt}
    \setlength{\belowcaptionskip}{0pt}
    \includegraphics[width=1\linewidth]{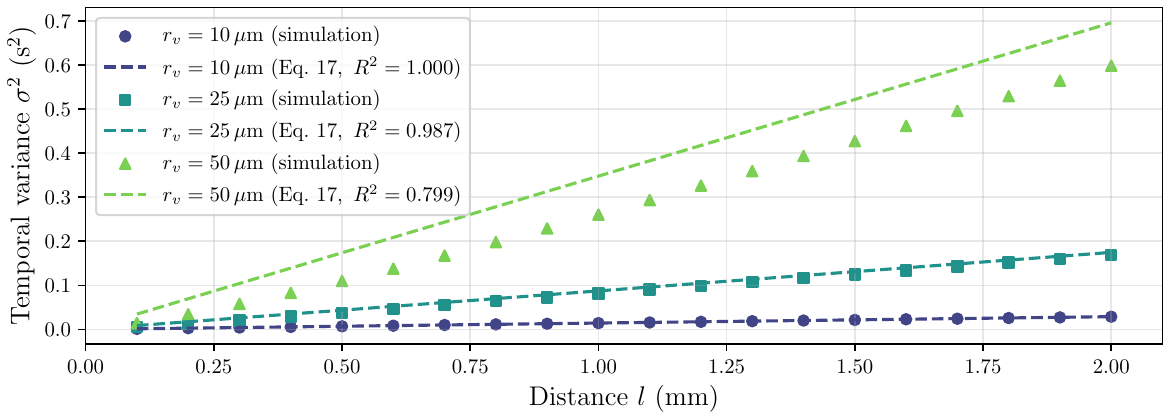}
    \caption{Temporal variance of the RX signal versus distance for different $r_v$ values. Results verify that our model works well in the lower $r_v$ limit.} 
    \label{fig:varianceapproxplotforr_v}
\end{figure}

\begin{figure}[t]
    \centering
    \setlength{\abovecaptionskip}{0pt}
    \setlength{\belowcaptionskip}{0pt}
    \includegraphics[width=1\linewidth]{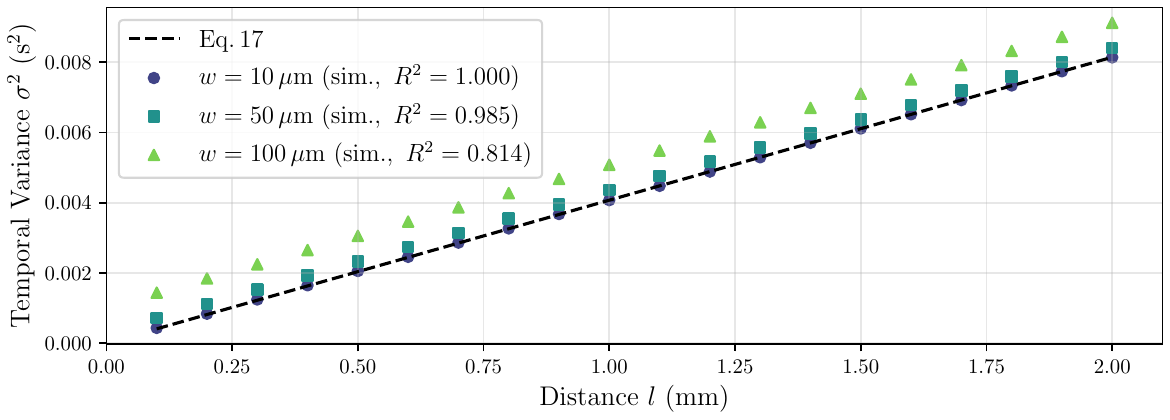}
    \caption{Temporal variance of the RX signal versus distance for different $w$ values. Results verify that our model works well in the small $w$ limit.}
    \label{fig:varianceapproxplotforw}
\end{figure}

We next evaluate the VALOR by considering the effect of the variables $v$, $r_v$, and $w$. 
As predicted by (\ref{varianceapprox}), the variance increases linearly with TX-RX distance, and the slope depends on the flow velocity. 
In Fig.~\ref{fig:varianceapproxplotforv}, simulation results (markers) are compared with the theoretical prediction (dashed lines). 
The match, with coefficients of determination $R^2 > 0.999$ across all tested velocities, confirms the accuracy of the Gaussian approximation in the capillary-scale regime. In Fig. \ref{fig:varianceapproxplotforr_v} and Fig. \ref{fig:varianceapproxplotforw}, the simulation results verify that our model works well on small $r_v$ and $w$ limits, as required by our derivation, while moving to the opposite extremes start to decrease the accuracy of the model. Please note that, with the parameters of a capillary vessel, the accuracy of our proposed model is high. Importantly, the variance--distance relation is robust to changes in velocity, meaning that once $v_{\textit{avg}}$ is known, the distance can be estimated reliably from a single receiver measurement without requiring the emission start time. This validates the VALOR as an accurate tool for TX localization in vessel-like molecular communication channels.
\begin{figure}[t]
    \centering
    \includegraphics[width=1\linewidth]{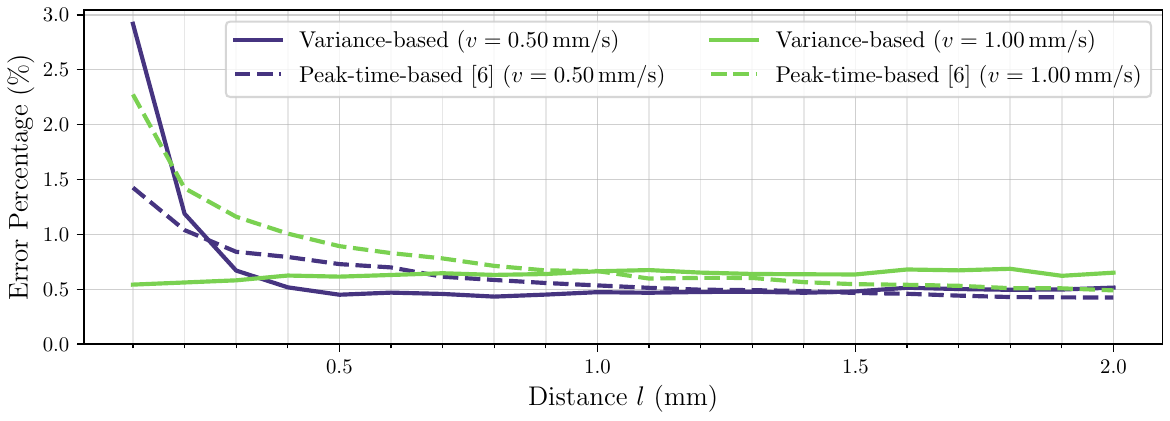}
    \caption{Percentage prediction error versus distance for variance-based and peak-time-based localization methods at different flow velocities. Solid lines show the variance-based estimator, while dashed lines show the peak-time method. Simulation results are averaged across replications. ($D = 300 \mu m^2/s, r_v = 5\mu m, w = 1\mu m$)}
    \label{fig:variancevspeak}
\end{figure}

In Fig.~\ref{fig:variancevspeak}, we compare the variance-based estimator with the peak-time method in \cite{VLE_localization}.  
Overall, the two approaches achieve comparable error levels, however \cite{VLE_localization} assumes the knowledge of the emission time while our method does not require it. Please note that knowing the emission time simplifies the problem significantly.
For the variance-based estimator, errors increase with flow velocity, reflecting that the validity condition in (\ref{condition1}) becomes less strict. 
At lower velocities, larger errors are observed at very short distances, where (\ref{condition2}) holds less strongly and statistical fluctuations dominate. 
Nevertheless, for all velocities the error quickly stabilizes once the TX-RX separation reaches the millimeter scale, remaining consistently below 1\%. 
This demonstrates that the variance-based method achieves accuracy similar to the peak-time approach, while requiring neither emission-time information nor multiple RXs.

\section{Conclusion and Future Directions}

In this letter, we introduced a variance-based method for TX localization in vessel-like molecular communication channels. 
By deriving a closed-form relation between the temporal variance of the 
received signal and the TX–RX distance, we obtained an estimator 
that operates with a single receiver and does not require knowledge of the 
emission time. Simulation results in capillary-scale environments confirmed 
that the VALOR achieves distance prediction errors on the order of 
1\%, with accuracy comparable to peak-time-based approaches that rely on 
additional assumptions.

Future work will extend this study from a single cylindrical vessel to 
connected vessel networks. In particular, applying the variance-based method 
to capillary networks with branching and varying geometries with different environment parameters will provide a 
more realistic understanding of its applicability in biological systems.



\bibliographystyle{IEEEtran}
\bibliography{references}

\end{document}